\documentclass[twocolumn,superscriptaddress,prb,floatfix,twoside,showpacs]{revtex4-1}
\pdfoutput=1
\usepackage{amsmath,graphics}

\IfFileExists{mtpro2.sty}{%

\usepackage[lite,subscriptcorrection]{mtpro2}
\linespread{1.04}
}{\relax}

\IfFileExists{microtype.sty}{\usepackage{microtype}}{\relax}

\usepackage{hyperref}
\hypersetup{
 pdfpagemode=UseNone,
 pdfstartview=FitH
}

\def\rmd{{\rm d}}
\def\rmi{{\rm i}}
\def\rme{{\rm e}}
\def\im{\mathop{\rm Im}\nolimits}

\def\figfrac{0.85}

\begin{document}

\title{Valence-band satellite in the ferromagnetic nickel: LDA+DMFT study
  with exact diagonalization}

\author{Jind\v{r}ich Koloren\v{c}}\email{kolorenc@fzu.cz}
\affiliation{Institut f\"ur Theoretische Physik, Universit\"at
  Hamburg, Jungiusstra\ss e 9, D-20355 Hamburg, Germany}
\affiliation{Institute of Physics, Academy of Sciences of
the Czech Republic, Na Slovance 2, CZ-182 21 Praha 8, Czech Republic}

\author{Alexander I. Poteryaev}
\affiliation{Institute of Metal Physics, Russian Academy of Sciences,
620990 Ekaterinburg, Russia}
\affiliation{Institute of Quantum Materials Science,
620107 Ekaterinburg, Russia}

\author{Alexander I. Lichtenstein}
\affiliation{Institut f\"ur Theoretische Physik, Universit\"at
  Hamburg, Jungiusstra\ss e 9, D-20355 Hamburg,
Germany}

\date{\today}

\begin{abstract}
The valence-band spectrum of the ferromagnetic nickel is calculated
using the LDA+DMFT method. The auxiliary impurity model emerging in
the course of the calculations is discretized
and solved with the exact diagonalization, or, more
precisely, with the Lanczos method. Particular emphasis is
given to spin dependence of the valence-band satellite that is
observed around 6 eV below the Fermi level. The calculated satellite
is strongly spin polarized in accord with experimental findings.
\end{abstract}

\pacs{ 71.20.Be, 71.15.$-$m, 75.30.$-$m}

\maketitle


\section{Introduction}

The electronic structure of transition metals has been intensively
studied for a number of decades. Notwithstanding, certain aspects of
the electron behavior in these materials are still not
completely understood. Comparison of experimental findings
with effective one-electron band theories have indicated that a more
thorough treatment of quantum many-body effects is necessary to
accurately describe the physical reality.

A prototypical metal displaying pronounced electron
correlations is the ferromagnetic nickel, where the
one-particle spectrum obtained using the local-density
approximation (LDA) to the density-functional theory (DFT) noticeably
departs from the spectra measured in photoemission
experiments. The calculated $3d$ bandwidth as well as the
exchange splitting are too large.\cite{eastman1978,eberhardt1980}
Moreover, the LDA completely misses the satellite feature located at
approximately 6 eV below the Fermi
level.\cite{fadley1968,baer1970,hufner1974} This satellite was
originally attributed to plasmon excitations,\cite{baer1970} but an
alternative view was soon proposed,\cite{hufner1975,kemeny1975}
according to which the satellite is a result of a correlated state of two
$3d$ holes localized in a single atom. The latter picture is supported
by the observed resonant enhancement of the satellite, during
which a second scattering channel involving $3p$ electrons and ending
in the same two-hole final state becomes
active.\cite{guillot1977,feldkamp1979,clauberg1981} The correspondence
between the satellite and the localized two-hole states can be explicitly
visualized in simplified finite-sized models that allow for an exact
many-body solution.\cite{victora1985,tanaka1992}

A more quantitative description of the electron correlations in nickel
can be achieved by incorporating a selfenergy into the LDA
or Hartree--Fock
bandstructure.\cite{penn1979,liebsch1979,igarashi1994} Usually, the
selfenergy is assumed local, that is, wave-vector independent. The
most sophisticated local selfenergy is provided by the
dynamical-mean-field theory (DMFT)\cite{georges1996} that maps the
problem of interacting lattice electrons onto an impurity model where
the interactions are restricted to a single lattice site. The combination
of LDA and DMFT (the so-called LDA+DMFT method) was applied to the
electronic structure of nickel several times in the past, using
different methods to solve the auxiliary impurity
model.\cite{lichtenstein2001a,katsnelson2002a,braun2006,grechnev2007,benea2012}
A reasonable
description was achieved employing the Hirsch--Fye quantum Monte Carlo (QMC)
method as the impurity solver.\cite{lichtenstein2001a} The QMC
methods have many merits. In particular, they are consistently accurate
regardless of the strength of correlations in the system. But they have
weaknesses too. The QMC calculations of the one-particle spectral
functions involve a numerical continuation from the imaginary
time to the real frequencies, a procedure with a limited resolution
especially at higher binding energies. Additionally, the QMC algorithm
used in Ref.~\onlinecite{lichtenstein2001a} is limited to a
diagonal-only Coulomb interaction. This truncation breaks a subset of
symmetries characterizing the full Coulomb operator, which can lead to
undesirable side effects.

In this paper we solve the auxiliary impurity model of the LDA+DMFT
by means of the Lanczos method. This strategy involves a
discretization of the impurity model which represents an obvious
limitation on the achievable accuracy. The
sources of errors in this approach are, however, very different from those in
the QMC method and the two impurity solvers can thus offer
complementary information. Using the Lanczos method, the
one-particle Green's function can be evaluated directly anywhere in
the complex plane without resorting to any extrapolation. It
is also straightforward to compare the full and truncated Coulomb
operators, and we make this comparison in Sec.~\ref{sec:results}.


\section{Method}
\label{sec:method}

We start from the bare electronic structure of Ni expressed in terms of a
tight-binding LMTO model\cite{andersen1975} containing $4s$, $3d$ and $4p$
electronic states. The one-electron Hamiltonian $\hat
H(\mathbf{k})$ is obtained as a solution of the local-density
approximation and the correlations beyond this approximation are
accounted for by a local selfenergy $\hat\Sigma$ acting in the
subspace of the $d$ orbitals. The selfenergy is spin polarized whereas
the Hamiltonian $\hat H(\mathbf{k})$ is spin independent. Taking $\hat
H(\mathbf{k})$ from the spin-polarized LDA is also possible,
although this route was found as less accurate
earlier.\cite{katsnelson2002a}

The selfenergy $\hat\Sigma$ is constructed with the aid of an impurity
model defined by a 
Hamiltonian $\hat H_{\rm imp}=\hat H^{(0)}_{\rm imp}+\hat U$ that
describes a single $d$ shell hybridized with a sea of auxiliary
conduction electrons. These auxiliary electrons, often referred to as
the bath, model the environment around the $d$ shell in the actual nickel
lattice. The Coulomb interaction $\hat U$ acts only among the $d$
orbitals and the one-particle part $\hat H^{(0)}_{\rm imp}$ has the form
\begin{align}
\label{eq:Himp0}
\hat H^{(0)}_{\rm imp}= 
& \sum_{m\sigma}\epsilon_{m\sigma}
 \hat d^{\dagger}_{m \sigma} \hat d_{m \sigma}
+ \sum_{km\sigma}
   \epsilon_{km\sigma} \hat c^{\dagger}_{km\sigma} \hat c_{km\sigma}
\nonumber\\[.2em]
&+ \sum_{km\sigma} V_{km\sigma}
  \Bigl( \hat d^{\dagger}_{m\sigma} \hat c_{km\sigma} +
         \hat c^{\dagger}_{km\sigma} \hat d_{m\sigma} \Bigr)\,,
\end{align}
where $\hat d^{\dagger}_{m \sigma}$ creates an electron in the
$d$~shell and $\hat c^{\dagger}_{km\sigma}$ creates a conduction
electron in the bath. The 
index $m$ runs over $e_g=\{x^2-y^2, z^2\}$ and $t_{2g}= \{xy, xz, yz\}$
states, and $\sigma\in\{\uparrow,\downarrow\}$ labels spin
projections. The hybridization parameters $V_{km\sigma}$ couple
only those impurity and bath levels that carry the same indices $m$
and $\sigma$, and hence the cubic symmetry and the electron spins are
preserved.

Provided we can solve the interacting impurity model, the sought for
selfenergy $\hat\Sigma$ is obtained as
\begin{subequations}
\label{eq:DMFT}
\begin{equation}
\hat\Sigma
=\hat G_{\rm imp}^{-1}\bigl[\hat H^{(0)}_{\rm imp}\bigr]
 - \hat G_{\rm imp}^{-1}\bigl[
\hat H_{\rm imp} 
\bigr]\,,
\end{equation}
where $\hat G_{\rm imp}[\hat H]$ represents the Green's function
matrix in the $d$-orbital subspace evaluated for a general impurity
Hamiltonian~$\hat H$. The matrix $\hat G_{\rm imp}\bigl[\hat
H^{(0)}_{\rm imp}\bigr]$, which we will denote as $\hat{\mathcal{G}}$
for short, is usually referred to as the bath Green's function.

So far, we have not specified how the parameters entering the Hamiltonian
$\hat H_{\rm imp}$ should be determined. The missing link
to the original lattice electrons is provided by a condition
\begin{equation}
\label{eq:DMFT_GG}
\hat G_{\rm imp}\bigl[\hat H_{\rm imp}\bigr]=
\hat G\bigl[\hat H(\mathbf{k}),\hat\Sigma\bigr]
\end{equation}
\end{subequations}
that equates $\hat G_{\rm imp}$ to the
local $d$-orbital Green's function $\hat G$ evaluated in the
lattice. The right-hand side of Eq.~\eqref{eq:DMFT_GG} can be
expressed as a momentum sum over the first Brillouin zone
\begin{equation}
\label{eq:GFloc_latt}
\hat G(z)=\frac1{N}\sum_k
 \bigl[(z+\mu)\hat I-\hat H(\mathbf{k})
 -\hat\Sigma(z)\bigr]^{-1}\,,
\end{equation}
where $\hat I$ stands for the identity operator and the chemical
potential $\mu$ is chosen such that the $4s$-$3d$-$4p$ space holds
ten electrons per Ni atom.

Equations~\eqref{eq:DMFT} define the dynamical-mean-field
approximation. They are iteratively solved for
$\hat\Sigma$ and $\hat H^{(0)}_{\rm imp}$ while $\hat
H(\mathbf{k})$ and $\hat U$ act as inputs. The most involved part of
this procedure is the solution of the multi-orbital impurity
model. A number of
approximations of varied accuracy have been used to find this solution
in the context of the DMFT. Here we discretize the impurity
Hamiltonian $\hat H_{\rm  imp}$ and then solve the resulting finite-sized 
cluster $\hat H_{\rm c}$ essentially exactly by means of the Lanczos
method. This strategy was successfully applied to the DMFT
equations for the repulsive\cite{caffarel1994,georges1996} and
attractive\cite{privitera2010} single-band Hubbard models as well as
for realistic multi-band problems.\cite{perroni2007}
The discretization $\hat H_{\rm  imp}\to\hat H_{\rm c}$ amounts to a
replacement of the infinite 
sums (integrals) over $k$ in Eq.~\eqref{eq:Himp0} with short finite sums. In
our particular case, the index $k$ takes only two values, that is, each
impurity orbital is connected to just two bath orbitals.


The parameters of the discretized Hamiltonian ($\epsilon_{m\sigma}$, 
$\epsilon_{km\sigma}$ and $V_{km\sigma}$) are expressed as functions
of $\hat H(\mathbf{k})$ and $\hat\Sigma$ with the aid of the
relation
\begin{equation}
\label{eq:G0_def}
\hat{\mathcal{G}}_{\rm c}^{-1}\equiv
\hat G_{\rm imp}^{-1}\bigl[\hat H^{(0)}_{\rm c}\bigr]
\approx\hat G^{-1}\bigl[\hat
H(\mathbf{k}),\hat\Sigma\bigr]+\hat\Sigma
=\hat{\mathcal{G}}^{-1},
\end{equation}
which is just a rearranged form of Eqs.~\eqref{eq:DMFT}. At this point
it is necessary to specify in what sense the discrete
bath Green's function $\hat{\mathcal{G}}_{\rm c}$ approximates the
continuous function $\hat{\mathcal{G}}$, that is, what is the precise
meaning of the symbol $\approx$ in Eq.~\eqref{eq:G0_def}. It has
become customary to minimize some distance between
$\hat{\mathcal{G}}_{\rm c}(z)$ and $\hat{\mathcal{G}}(z)$ defined on
the Matsubara frequencies $z=\rmi\omega_n$. A particularly convenient
choice is a least-squares
fit,\cite{caffarel1994,georges1996,perroni2007,privitera2010}
for instance
\begin{equation}
\label{eq:fit}
\min_{\substack{\epsilon_{m\sigma},\epsilon_{km\sigma}\\[.2em]
V_{km\sigma}}}
\sum_n\biggl|
\frac1{\mathcal{G}^{\rm c}_{m\sigma}(\rmi\omega_n)}-
\frac1{\mathcal{G}_{m\sigma}(\rmi\omega_n)}
\biggr|^2
\end{equation}
for each $m$ and $\sigma$.
For reasons that will be discussed later, we do 
not follow this fitting route but adopt an
alternative approach instead. We obtain the parameters of the
discretized Hamiltonian from the requirement of coincidence
of the high-frequency asymptotics of  $\hat{\mathcal{G}}_{\rm c}(z)$
and $\hat{\mathcal{G}}(z)$.\cite{si1994,georges1996}

The cluster Green's function $\hat{\mathcal{G}}_{\rm c}(z)$ can be
written in an explicit form\cite{hewson_kondo_problem}
\begin{equation}
\label{eq:G0_matsub}
\mathcal{G}^{\rm c}_{m\sigma}(z)=\biggl(z-\epsilon_{m\sigma}
  -\sum_k\frac{V_{km\sigma}^2}{z-\epsilon_{km\sigma}}\biggr)^{-1}
\end{equation}
whose expansion in powers of $1/z$ reads as
\begin{multline}
\label{eq:G0_expansion_discr}
\mathcal{G}^{\rm c}_{m\sigma}(z)
=\frac1z
+\frac{\epsilon_{m\sigma}}{z^2}
+\frac{\epsilon_{m\sigma}^2+\sum_k V_{km\sigma}^2}{z^3}\\[.2em]
+\frac{\epsilon_{m\sigma}^3+\sum_k V_{km\sigma}^2
  (\epsilon_{km\sigma}+2\epsilon_{m\sigma})}{z^4}
+\cdots
\end{multline}
The continuous Green's function $\hat{\mathcal{G}}(z)$ can be expressed
in terms of the density of states $g(z)$, and the coefficients of
the expansion in powers of $1/z$ are then given as moments of this
density of states,
\begin{equation}
\label{eq:G0_expansion_cont}
\mathcal{G}_{m\sigma}(z)
  =\int\frac{g_{m\sigma}(\epsilon)}{z-\epsilon}\,\rmd\epsilon
=\sum_{n=1}^{\infty}\frac1{z^n}
  \underbrace{%
  \int\epsilon^{n-1} g_{m\sigma}(\epsilon)\,\rmd\epsilon}_{%
  \displaystyle M_{n-1}}\,.
\end{equation}
With two bath orbitals per each impurity $d$ orbital we have five
parameters in $H^{(0)}_{\rm c}$ that carry the same indices $m$ and
$\sigma$, and thus we can match Eqs.~\eqref{eq:G0_expansion_discr}
and~\eqref{eq:G0_expansion_cont} up to $1/z^6$.

We are mostly interested in the ground-state properties and in the
one-particle spectrum in, say, the first 10 eV below the Fermi
level. In the course of our calculations we observed that the
discretization procedure defined by
Eqs.~\eqref{eq:G0_def}--\eqref{eq:G0_expansion_cont} often placed
some of the bath energies $\epsilon_{km\sigma}$ quite high above the
Fermi level far outside the energy window of our interest. That
by itself would not be an issue if it did not lead to an unphysical
stabilization of a non-magnetic solution. (See Appendix~\ref{app:B} for
an illustration and further discussion of the effect). In order to
suppress this undesirable behavior, we modify the definition of the
moments $M_n$ to
\begin{equation}
\label{eq:moments_cutoff}
M_{n}=\frac{\displaystyle%
\int_{\epsilon_l}^{\epsilon_u}\epsilon^{n}
   g_{m\sigma}(\epsilon)\,\rmd\epsilon}{\displaystyle%
\int_{\epsilon_l}^{\epsilon_u}
   g_{m\sigma}(\epsilon)\,\rmd\epsilon}\,.
\end{equation}
The lower cutoff is a purely technical matter;
it is set to $\epsilon_l=-9$ eV, that is, below the $4s$ band. The
upper cutoff avoids the unphysical solution by not allowing the bath
orbitals to drift to high energies. The results presented
in Sec.~\ref{sec:results} were obtained with $\epsilon_u=2$~eV.
The possibility to straightforwardly prevent the non-magnetic state
with the aid of the upper cutoff $\epsilon_u$ is the main reason
why we opted for the bath discretization by means of the $1/z$
expansion instead of the more frequently employed fitting on the
Matsubara axis. We have not succeeded in finding a suitable
modification of the fitting function, Eq.~\eqref{eq:fit}, that would
reliably eliminate the non-magnetic solution.


The last component of the cluster Hamiltonian $\hat
H_{\rm c}$ is the Coulomb interaction in the $d$ shell. We use the
spherically symmetric form
\begin{multline}
\label{eq:Coulomb_potential}
\hat U= \frac{1}{2}
\sum_{\substack {m m' m''\\  m''' \sigma \sigma'}} 
  U_{m m' m'' m'''} \hat d^{\dagger}_{m\sigma} \hat d^{\dagger}_{m' \sigma'}
  \hat d_{m'''\sigma'} \hat d_{m'' \sigma} \\[-1.2em]
-U_{\rm H}\sum_{m\sigma} \hat d^{\dagger}_{m \sigma} \hat d_{m \sigma}\,,
\end{multline}
where the matrix $U_{m m' m'' m'''}$ is
parametrized by the Slater integrals $F_0=2$ eV, $F_2=8.2$ eV and
$F_4=5.2$ eV. These numerical values correspond to Coulomb
$U=2$ eV and exchange $J=0.95$ eV.
The contribution to Eq.~\eqref{eq:Coulomb_potential} proportional
to $U_{\rm H}$ represents a rigid shift of the impurity levels
downward, $\epsilon_{m\sigma}\to \epsilon_{m\sigma}-U_{\rm H}$, and
accounts for the fact that the $d$--$d$
Coulomb interactions are already partially included in the LDA Hamiltonian
$\hat H(\mathbf{k})$ in the form of a static mean field. Several
formulas have been proposed to express the Hartree-like
double-counting potential $U_{\rm H}$ in terms of the occupation
numbers of the $d$ orbitals,\cite{anisimov1991,czyzyk1994,solovyev1994}
but we treat $U_{\rm H}$ 
as a free parameter similarly to Ref.~\onlinecite{karolak2010}, since
neither of the standard choices leads to satisfactory results.

The need for ``undressing'' the LDA quasiparticles is one of the
reasons why we prefer to build the many-body description on the top of
the spin-restricted LDA bandstructure. If we started from polarized
bands, the Hartree potential $U_{\rm H}$ would be polarized too, which
would introduce an extra complexity to the problem. The double
counting would have to be spin dependent also in the LDA+DMFT
implementations that take into account the feedback of the selfenergy
on $\hat H(\mathbf{k})$.\cite{shick2009,haule2010}

Once the cluster Hamiltonian $\hat H_{\rm c}$ is fully specified,
the one-particle Green's function
$\hat G_{\rm c}\equiv\hat G_{\rm imp}[\hat H_{\rm c}]$ for individual $d$
orbitals can be calculated. We
employ the band Lanczos method\cite{ruhe1979,meyer1989} that allows
for a simultaneous evaluation of all relevant matrix elements at
once. Off-diagonal elements are directly accessible too, although this
functionality is not used in the application at hand. For the
purpose of the Lanczos method, $\hat G_{\rm c}$ is decomposed 
in the following form \cite{capone2007}
\begin{equation}
G^{\rm c}_{m\sigma}(z)=\frac1Z
\bigl[G_{m\sigma}^{>}(z)+G_{m\sigma}^{<}(z)\bigr]\,,
\end{equation}
where the two parts are
\begin{align*}
G_{m\sigma}^{>}(z)&=\sum_{\alpha}\rme^{-\beta E_{\alpha}}
 \langle\alpha|\hat d_{m\sigma}
 \bigl(z+E_{\alpha}-\hat H_{\rm c}\bigr)^{-1}
 \hat d^{\dagger}_{m\sigma}|\alpha\rangle\,, \\
G_{m\sigma}^{<}(z)&=\sum_{\alpha}\rme^{-\beta E_{\alpha}}
 \langle\alpha|\hat d^{\dagger}_{m\sigma}
 \bigl(z-E_{\alpha}+\hat H_{\rm c}\bigr)^{-1}
 \hat d_{m\sigma}|\alpha\rangle\,.
\end{align*}
The sums over the many-body eigenstates $|\alpha\rangle$,
$\hat H_{\rm c}|\alpha\rangle=E_{\alpha}|\alpha\rangle$, represent
grandcanonical averages with the chemical potential equal zero, and
$Z=\sum_{\alpha}\rme^{-\beta E_{\alpha}}$ stands for the corresponding
partition function. The calculations are performed at low temperature
$k_{\rm B}T=1/\beta=0.01$ eV so that only the ground state
contributes to the sum over $\alpha$ most of the time. The
eigenstate-eigenvalue pairs including all their degeneracies are found
using the implicitly restarted Lanczos method as implemented in the
ARPACK software package.\cite{arpack}

\section{Results and discussion}
\label{sec:results}

\subsection{Ground-state properties}

\begin{figure}[t]
\resizebox{\figfrac\linewidth}{!}{\includegraphics{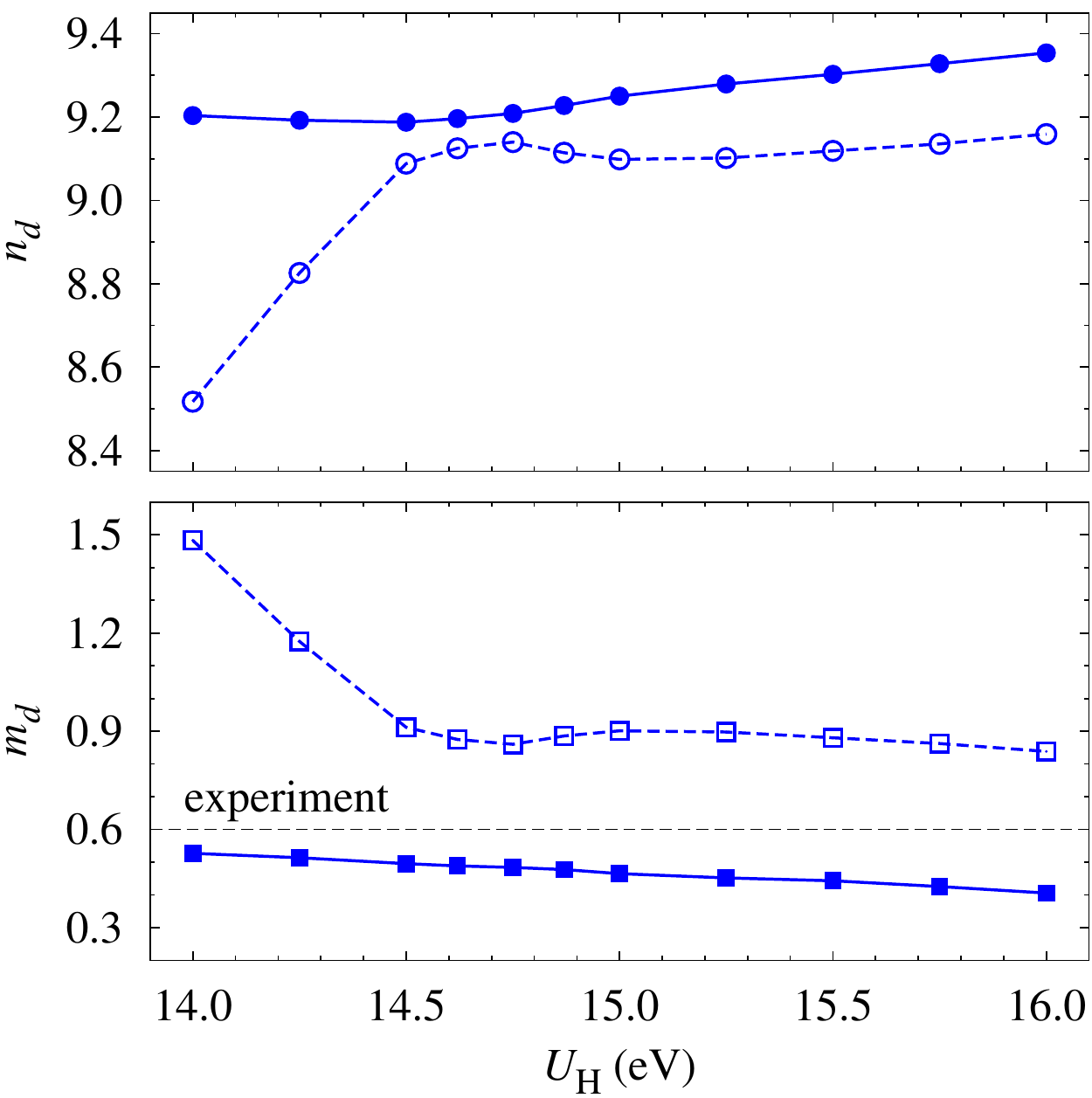}}
\caption{\label{fig:nd_and_pol}
  (color online) The occupation of the $d$ orbitals $n_d$ (top) and
  the spin polarization $m_{d}$ (bottom) plotted as
  functions of the double-counting potential $U_{\rm H}$. Empty
  symbols correspond to the cluster Green's function
  $\hat G_{\rm c}$, full symbols to the lattice Green's function
  $\hat G$.}
\end{figure}

First we examine selected characteristics of the ground state and use them
to estimate the double-counting
potential $U_{\rm H}$. Figure~\ref{fig:nd_and_pol} shows the number of
electrons in the $d$ orbitals $n_d=n_{d\uparrow}+n_{d\downarrow}$ and
the spin polarization of these
orbitals $m_d=n_{d\uparrow}-n_{d\downarrow}$. The data calculated in the
lattice and in the discretized impurity model are plotted side by 
side. They differ despite the DMFT iterations being
converged in the sense that the cluster Hamiltonian $H_{\rm c}$
no longer changed in the successive steps. The differences would
vanish if we solved the full continuous impurity model, since
Eq.~\eqref{eq:DMFT_GG} would be exactly fulfilled in that case.

It turns out that $n_d$ and $m_d$ depend only weakly on
the double-counting potential $U_{\rm H}$ when the latter is larger
than approximately $14.5$ eV. Below $14.5$ eV the
trend changes and the cluster quantities depart 
substantially from their lattice counterparts. Based on this
observation we consider $U_{\rm H}$ below $14.5$ eV as
inappropriate. We note in passing that the double counting in the
so-called fully localized limit\cite{czyzyk1994,solovyev1994}
$U_{\rm H}^{\rm (FLL)}=U(n_d-1/2)-J(n_d-1)/2$ equals
$13.2$ eV for $n_d=9$ and it is thus more than 1 eV too
small to be applicable in our case. The so-called around mean-field
form\cite{anisimov1991} of $U_{\rm H}$ provides an even
smaller value.

The experimentally determined magnetization of the fcc nickel is
approximately 0.6 $\mu_{\rm B}$ per atom.\cite{danan1968} Our
calculations slightly underestimate this quantity even though the
cluster solution, from which the spin-dependent selfenergy is
extracted, displays the maximal polarization characterized by
$m_d=5-n_{d\downarrow}$.

The number of $d$ electrons cannot be unambiguously defined in a solid
and as such it does not represent a particularly useful measure of
quality of our ground state. The $d$-band filling in nickel is often
estimated as 9.4 per atom based on the measured magnetic moment and
the assumption of the maximal $d$-shell polarization,\cite{mott1964}
but reliability of this estimate is limited.

\subsection{Valence-band spectrum}

\begin{figure}[t]
\resizebox{\figfrac\linewidth}{!}{\includegraphics{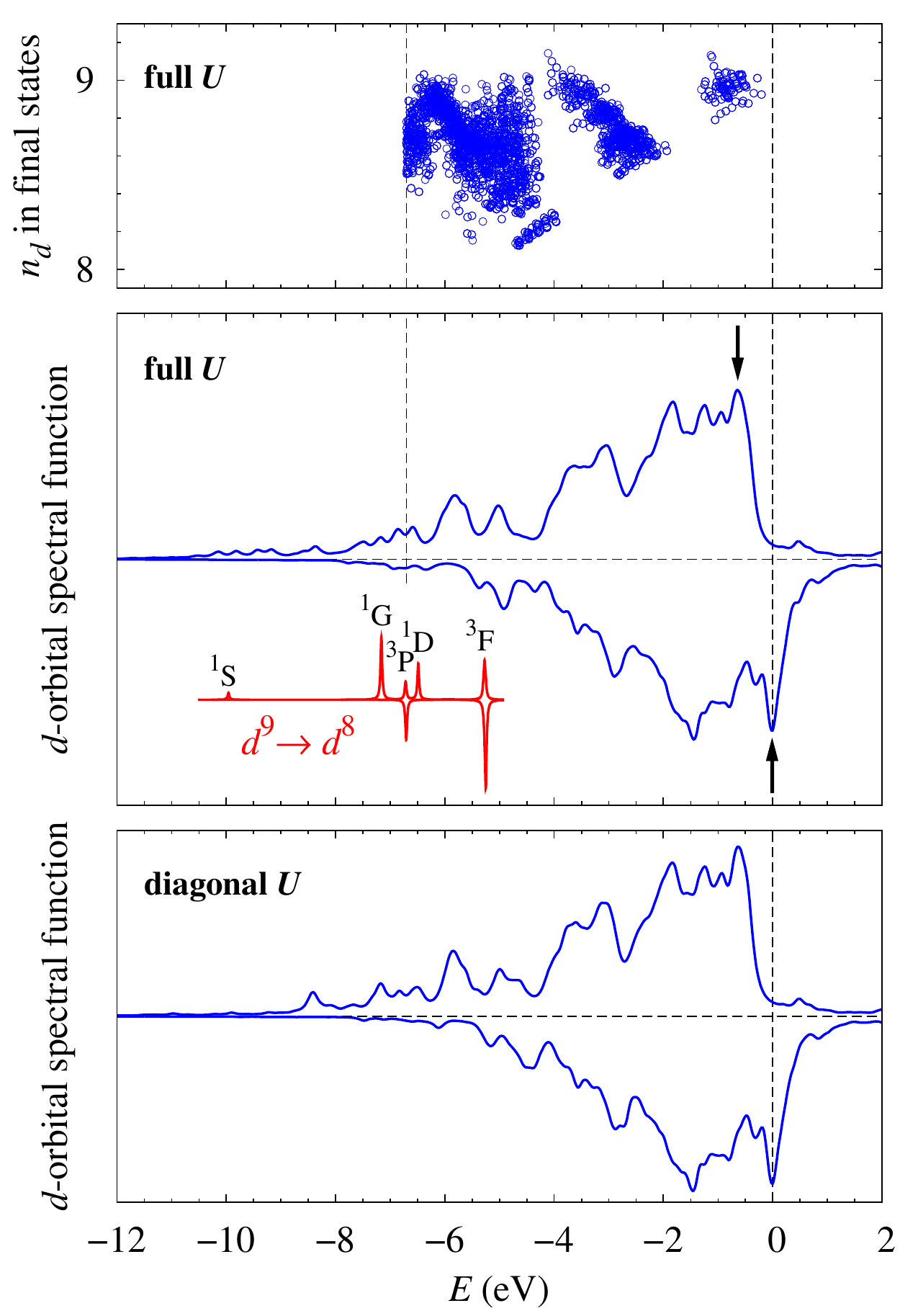}}
\caption{\label{fig:dos_DMFT}
  (color online) Spin-resolved $d$-orbital spectral function of bulk Ni
  obtained with $U_{\rm H}=15$ eV. The middle panel corresponds to
  the full Coulomb vertex, the bottom panel to the truncated
  vertex. The atomic $d^9 \to d^8$ transitions are displayed at an
  arbitrary scale and position in the middle panel for comparison with
  the shape of the
  satellite. The top panel shows the $d$-orbital occupation in the
  first 5000 many-body final states corresponding to the cluster
  Hamiltonian with the full vertex.}
\end{figure}

We find that the one-particle spectra corresponding to the
double-counting potential $U_{\rm H}$ in the range
$15.0\pm 0.5$ eV are only barely distinguishable.
Figure~\ref{fig:dos_DMFT} shows the $d$-orbital spectral function
$\im\sum_m\bigl[G_{m\sigma}(E-\rmi 0)\bigr]/\pi$ for $U_{\rm H}=15$
eV. The calculated spectrum is relatively disappointing: the
width of the main band ($\approx 4$ eV) as well as the exchange splitting
are nearly identical to those obtained with the spin-polarized LDA,
and thus share the same poor agreement with experiments. The
symmetry-resolved exchange splitting at the Fermi level is given
directly by the selfenergy and reads as
\begin{subequations}
\begin{align}
\Sigma_{e_g\uparrow}(E_{\rm F})
  -\Sigma_{e_g\downarrow}(E_{\rm F})&\approx 0.3 \text{ eV,} \\
\Sigma_{t_{2g}\uparrow}(E_{\rm F})
  -\Sigma_{t_{2g}\downarrow}(E_{\rm F})&\approx 0.8 \text{ eV.}
\end{align}
\end{subequations}
The $d$ states near the Fermi level have predominantly the $t_{2g}$
character, which results in the apparent exchange splitting of $0.6$~eV
that is visible as a shift between the top of the valence bands for the
minority and majority spins (indicated with arrows in
Fig.~\ref{fig:dos_DMFT}). Appendix~\ref{app:A} indicates that the band
width and the exchange splitting could possibly be improved if the
impurity model was discretized using Eq.~\eqref{eq:fit} instead of
Eqs.~\eqref{eq:G0_expansion_discr} and ~\eqref{eq:G0_expansion_cont},
but, as discussed in Sec.~\ref{sec:method}, that approach is not
sufficiently reliable.

We identify the spectral features below $-4.5$ eV as the ``6 eV
satellite''. It is strongly spin polarized in agreement with
spin-resolved photoemission experiments.\cite{see1995} In our
calculations, the energy-integrated spectral weight is about three
times larger for the majority spins than for the minority spins.
Furthermore, the minority-spin states are located at reduced binding
energies, which was also observed experimentally.\cite{kakizaki1997}
The calculated characteristics of the satellite corroborate its
explanation based on transitions from the spin-polarized $d^9$ initial
state to the $d^8$ final states. An illustration of such atomic
spectral lines is added to Fig.~\ref{fig:dos_DMFT} for comparison. The
singlet final states ${}^1$D, ${}^1$G and ${}^1$S exhibit a complete
majority-spin polarization and lie deeper, the triplet states ${}^3$F
and ${}^3$P carry a partial polarization in the opposite direction and
lie shallower.


This simplified description of the satellite should not be taken too
literally, however, at least not within our computational scheme. We have
calculated the $d$-orbital occupation $n_d$ corresponding to the
final states in our discretized impurity model, the results
are aligned with the lattice spectral function in
Fig.~\ref{fig:dos_DMFT}. Although $n_d$ indeed decreases as
the binding energy increases, it is still considerably larger than eight
in the satellite region where contributions from states with $n_d\geq
8.5$ are not an exception. This enhancement of $n_d$ is due to
impurity--bath hybridization as discussed at the end of
Appendix~\ref{app:B}. It is possible that $n_d$ is somewhat overestimated as a
result of the sparse discretization of the bath.

As mentioned earlier, our calculations are rather insensitive
to a choice of the potential $U_{\rm H}$ as long as it
exceeds a threshold of approximately 14.5 eV. For smaller $U_{\rm H}$
the impurity orbitals in the cluster start to depopulate, which is
accompanied by an increased intensity of the satellite. This result
is in accord with experiments on alloys of Ni with
electropositive metals.\cite{fuggle1981,fuggle1983}

Finally, we compare spectral functions calculated with
two versions of the Coulomb operator: the full spherically symmetric
vertex discussed so far, and the diagonal-only vertex employed in
the Hirsch--Fye QMC method.\cite{lichtenstein2001a}
Figure~\ref{fig:dos_DMFT} shows that the simplification of the interaction
has virtually no effect on the main $d$ bands. The satellite,
on the other hand, is visibly modified. The tail of the majority-spin spectrum
does not extend as deep as with the full interaction, and the
minority-spin satellite is shifted to smaller binding energies. It is
possible that this slight shift in conjunction with the low
resolution of the maximum-entropy method leads to a merger of the
satellite with the main band for the minority spins, resulting in a fully
spin-polarized satellite reported in Ref.~\onlinecite{lichtenstein2001a}.

\section{Conclusions}

We have investigated the valence-band spectra of the ferromagnetic
nickel within the LDA+DMFT framework. The auxiliary impurity model was
discretized and then solved using the Lanczos method. The valence-band
satellite and its spin dependence was reproduced in good agreement
with spin-resolved photoemission experiments. The
many-body renormalization of the $3d$ bands
as well as the exchange splitting are found to be sensitive to the details
of the bath discretization, which indicates that ten orbitals are
probably not enough to approximate the bath to a satisfactory
accuracy. The diagonalization method as employed in this paper is
adequate for recovering features of atomic origin located at high
binding energies but it is apparently too crude to capture the
expected modification of the Fermi-liquid parameters at low binding energies.

\begin{acknowledgments}
Financial support by the Deutsche Forschungsgemeinschaft through FOR
1346 is gratefully acknowledged. J.~K. acknowledges financial support
by the Alexander von Humboldt Foundation. A.~P. acknowledges the
Russian Foundation for Basic Research (Projects Nos. 10-02-00046,
10-02-91003, and 11-02-01443).
\end{acknowledgments}

\appendix
\section{Comments on the bath discretization}
\label{app:A}
\begin{figure}[t]
\resizebox{\figfrac\linewidth}{!}{\includegraphics{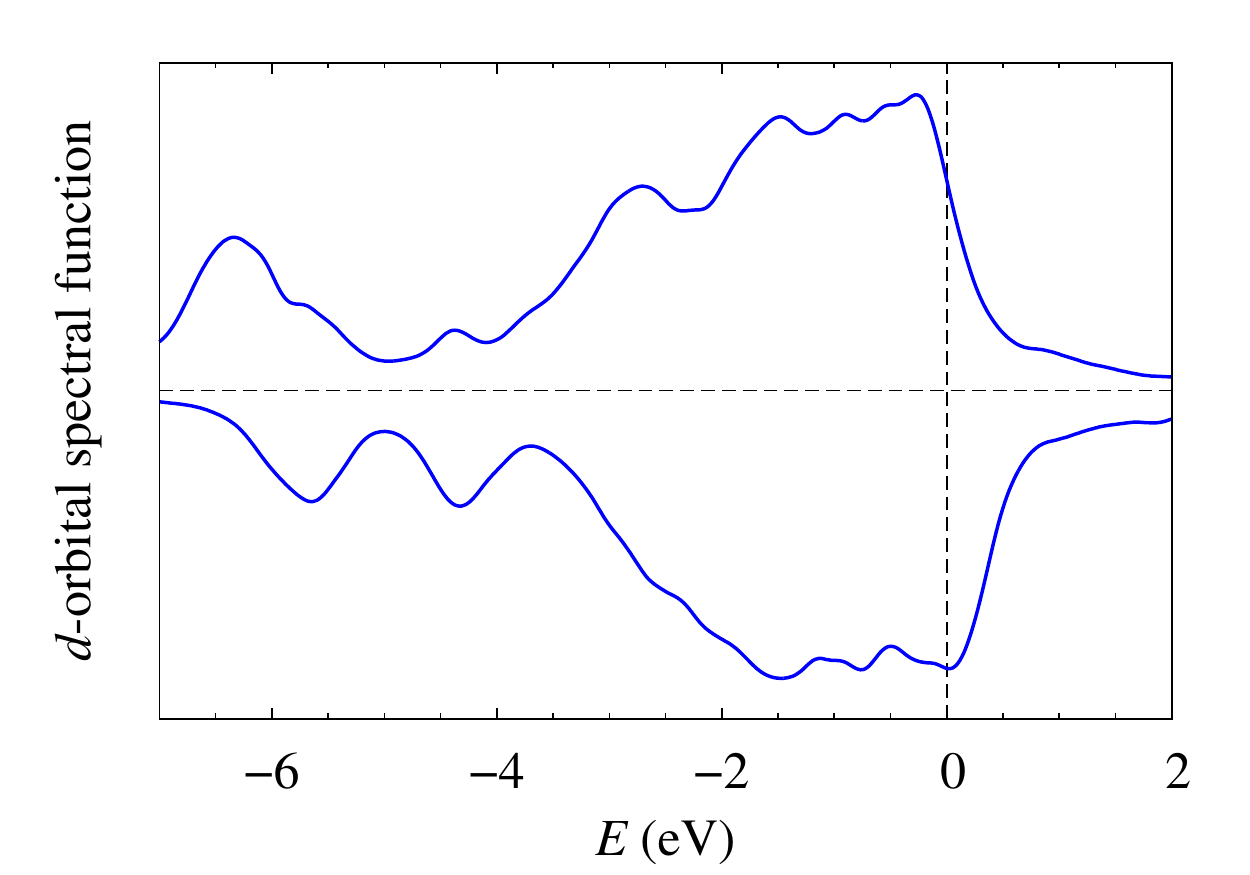}}
\caption{\label{fig:DOS_matsub}%
  (color online) Spin-resolved $d$-orbital spectral function of bulk Ni
  that was obtained with the bath discretized using the
  least-squares fit at the Matsubara
  frequencies (five orbitals in the bath).}
\end{figure}

In this appendix we return to the bath discretization using the
least-squares fit of the bath Green's function with the functional form of
Eq.~\eqref{eq:G0_matsub} at the Matsubara frequencies. In the case of
five orbitals in the bath, the self-consistent LDA+DMFT calculations
are very stable and the resulting spectral function is shown in the
Fig.~\ref{fig:DOS_matsub}. Although the satellite at high binding
energies is not described as well as previously
(Fig.~\ref{fig:dos_DMFT}), the renormalization of the main valence band
and the exchange splitting come out as more reasonable. The width
of the valence band is approximately 3 eV and the exchange spitting
is about 0.3 eV, both of which are close to the photoemission
experiments.\cite{eastman1978,eberhardt1980} The fitting at the Matsubara
frequencies is more sensitive to the behavior near the Fermi level
than the method of moments, Eqs.~\eqref{eq:G0_expansion_discr}
and~\eqref{eq:G0_expansion_cont}, and hence it gives a finer control over
the low-energy spectral features.
Unfortunately, due to the effects discussed in the next appendix, it is
not easy to converge the calculations to the correct state for
ten orbitals in the bath. Nevertheless, we believe that it will
eventually be
possible to find an optimal way of fitting the bath
Green's function that would lead to a reasonable description of both
low- and high-energy parts of the spectrum simultaneously.

\section{Magnetism in the finite cluster}
\label{app:B}

Here we take a closer look at the issues that necessitated
the introduction of the cutoffs in Eq.~\eqref{eq:moments_cutoff}. To
illustrate the problem, we use a simpler impurity model than that
employed in Sec.~\ref{sec:results}---we reduce the cluster to contain
only one bath orbital per each impurity orbital and we assume spherical
instead of cubic symmetry. Furthermore, we implement the Hamiltonian
parameters used in Ref.~\onlinecite{tanaka1992}, which
gives us the opportunity to relate our calculations to this earlier
study of electron correlations in nickel. The Slater integrals are
$F_0=3.5$ eV, $F_2=9.79$ eV and $F_4=6.08$ eV, and the impurity--bath
hopping is $V_{km\sigma}=0.7$ eV. The bath-level position
$\epsilon_{km\sigma}\equiv\epsilon_{\rm b}$ is treated as a free parameter
and the double-counting potential $U_{\rm H}$ is determined such that
there are always nine electrons in the impurity $d$ orbitals. The
temperature is $k_{\rm B}T=0.01$ eV as before.

\begin{figure}[t]
\resizebox{\figfrac\linewidth}{!}{\includegraphics{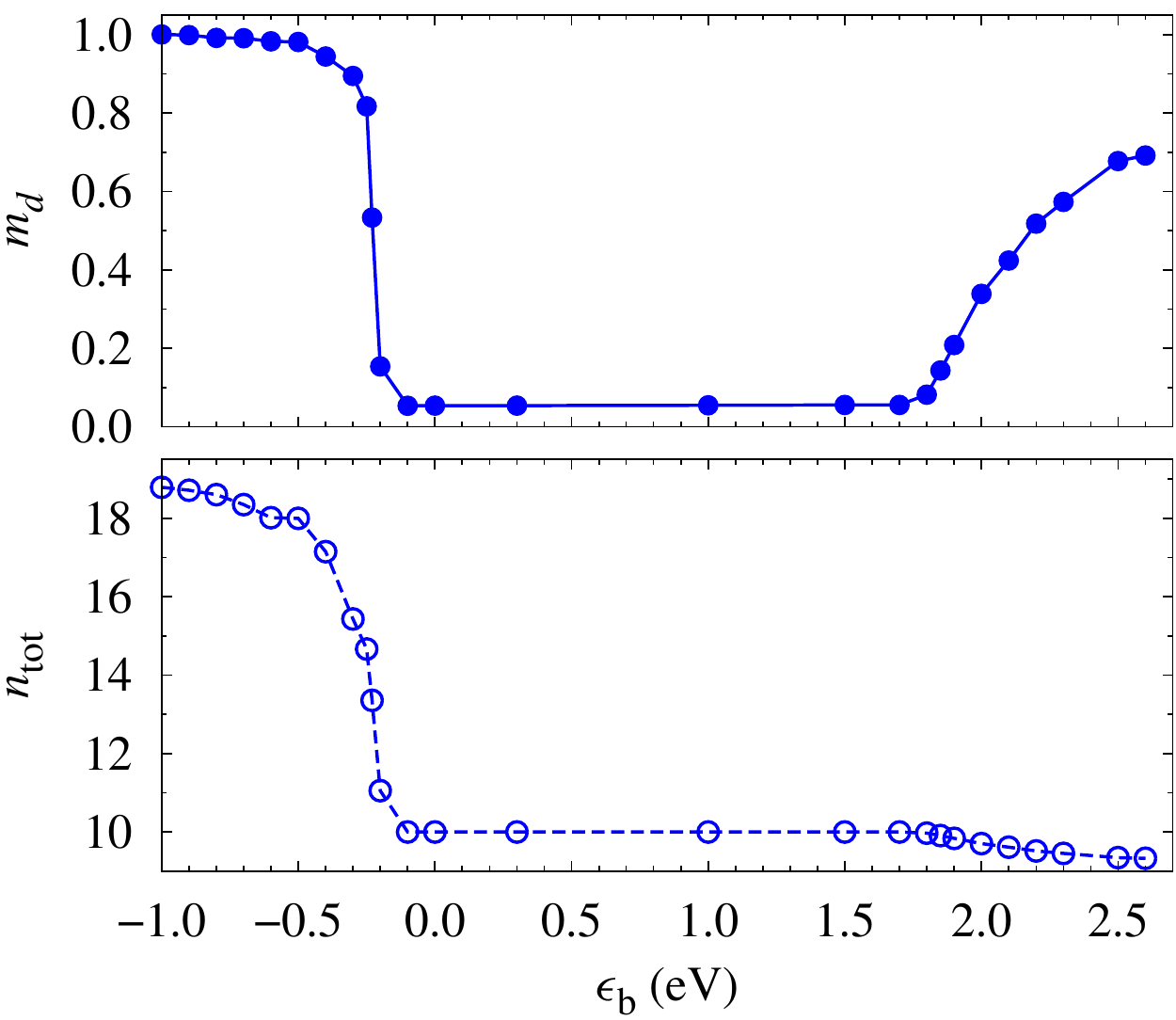}}
\caption{\label{fig:bath_sweep}%
  (color online) Spin polarization of the $d$ shell $m_d$  (top) and
  the total number of electrons in the whole cluster $n_{\rm tot}$
  (bottom) as functions of the bath-level position $\epsilon_{\rm b}$.}
\end{figure}

The spin polarization of the impurity orbitals is induced by a small
magnetic field $B$ coupled to the impurity spins. The coupling is
described by an extra term in the cluster Hamiltonian,
\begin{equation}
\hat H^{(B)}_{\rm c}=\frac{B}{2}\sum_m
 \left(\hat d^{\dagger}_{m\uparrow}\hat d_{m\uparrow}
 -\hat d^{\dagger}_{m\downarrow}\hat d_{m\downarrow} \right)\,.
\end{equation}
The resulting polarization is plotted in Fig.~\ref{fig:bath_sweep} as a
function of the bath position $\epsilon_{\rm b}$ scanned across the Fermi
level. The total number of electrons in the cluster is shown as well.


When the bath orbitals are sufficiently deep below the Fermi level, the
bath is nearly full and a local magnetic moment is formed on the
impurity.
As the bath orbitals move up toward the Fermi level, the bath relatively
quickly depopulates until it holds only a single electron. This
electron, together with the other nine sitting in the impurity orbitals,
forms a non-magnetic $d^{10}$ closed shell.
This state then remains stable
even when the bath is raised relatively high above the Fermi
level. Only for $\epsilon_{\rm b}>2.4 V_{km\sigma}\approx 1.7$ eV the
bath starts 
releasing the last electron and a magnetic ground state is
restored. The larger cluster corresponding to our actual DMFT
calculations shows an analogous behavior, only the non-magnetic
solution occurs for 20 electrons in the cluster as there is an extra
fully occupied shell of bath orbitals located deeper
below the Fermi level.

\begin{figure}[t]
\resizebox{\figfrac\linewidth}{!}{\includegraphics{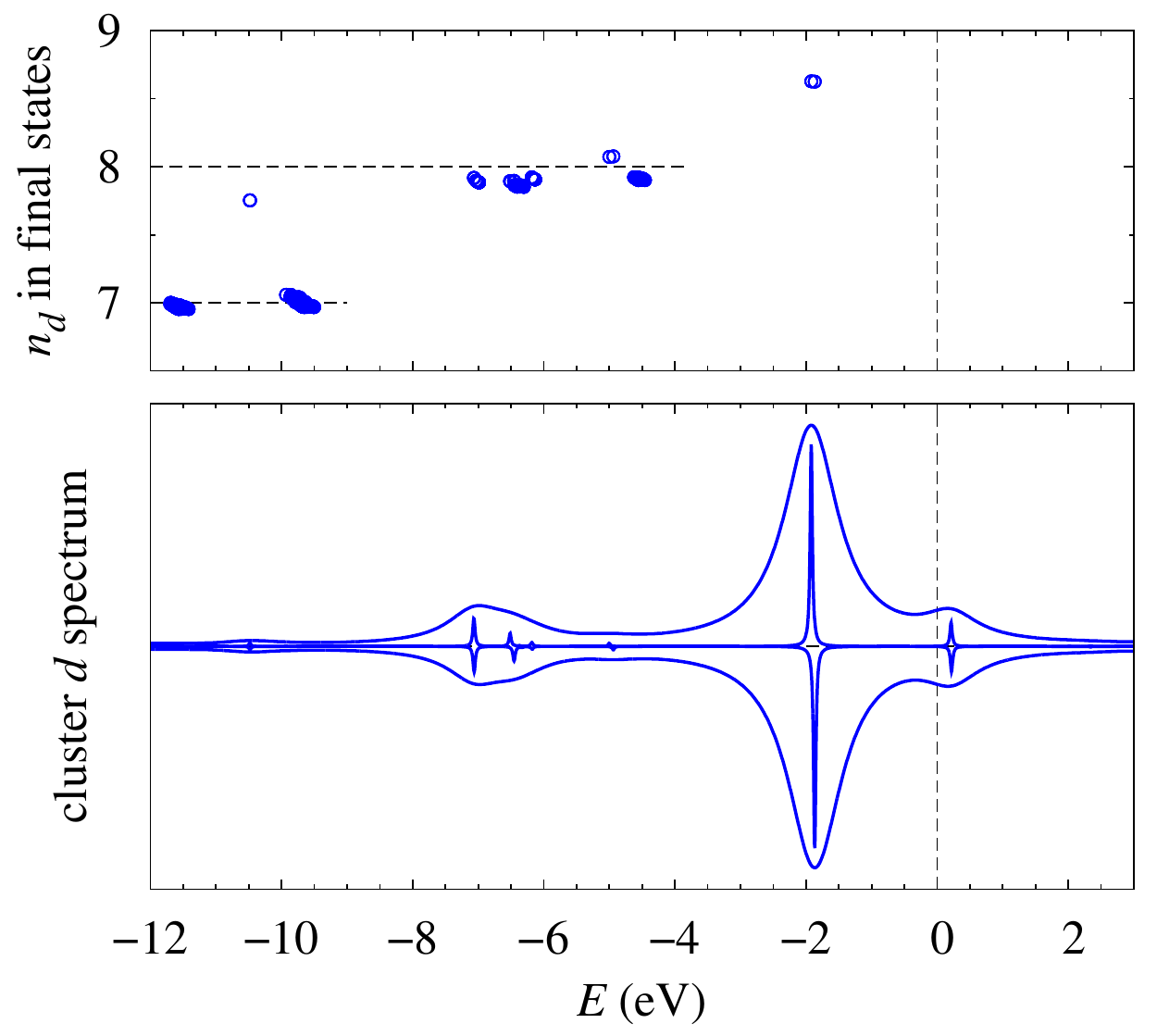}}
\caption{\label{fig:TJS_spectrum}%
  (color online) Spin-resolved $d$-orbital spectral function
  of a cluster with parameters taken from
  Ref.~\onlinecite{tanaka1992} and $\epsilon_{\rm b}=0$ eV (lower
  panel). The ``envelope'' is 
  calculated with a large Lorentz broadening of 0.5 eV. The
  top panel shows the $d$-orbital occupation in the final states.}
\end{figure}

It is clear that the non-magnetic solution does not correctly describe
the $d$ shell and its environment in the ferromagnetic
nickel. Elevating the bath orbitals high above the Fermi level in order to
support a magnetic ground state does not look as very plausible.
This leaves us with the configuration where the bath states are
nearly fully occupied and hence they model the nearly full $d$ orbitals of
the nickel atoms surrounding the ``impurity'' site. To prevent the cluster
Hamiltonian to enter the non-realistic regimes in the course of the
DMFT iterations, we have introduced the upper cutoff $\epsilon_u$ in
the integrals in Eq.~\eqref{eq:moments_cutoff}. This cutoff does not
allow the bath orbitals to drift too high and to lock into the
non-magnetic solution.

\begin{figure}[b]
\resizebox{\figfrac\linewidth}{!}{\includegraphics{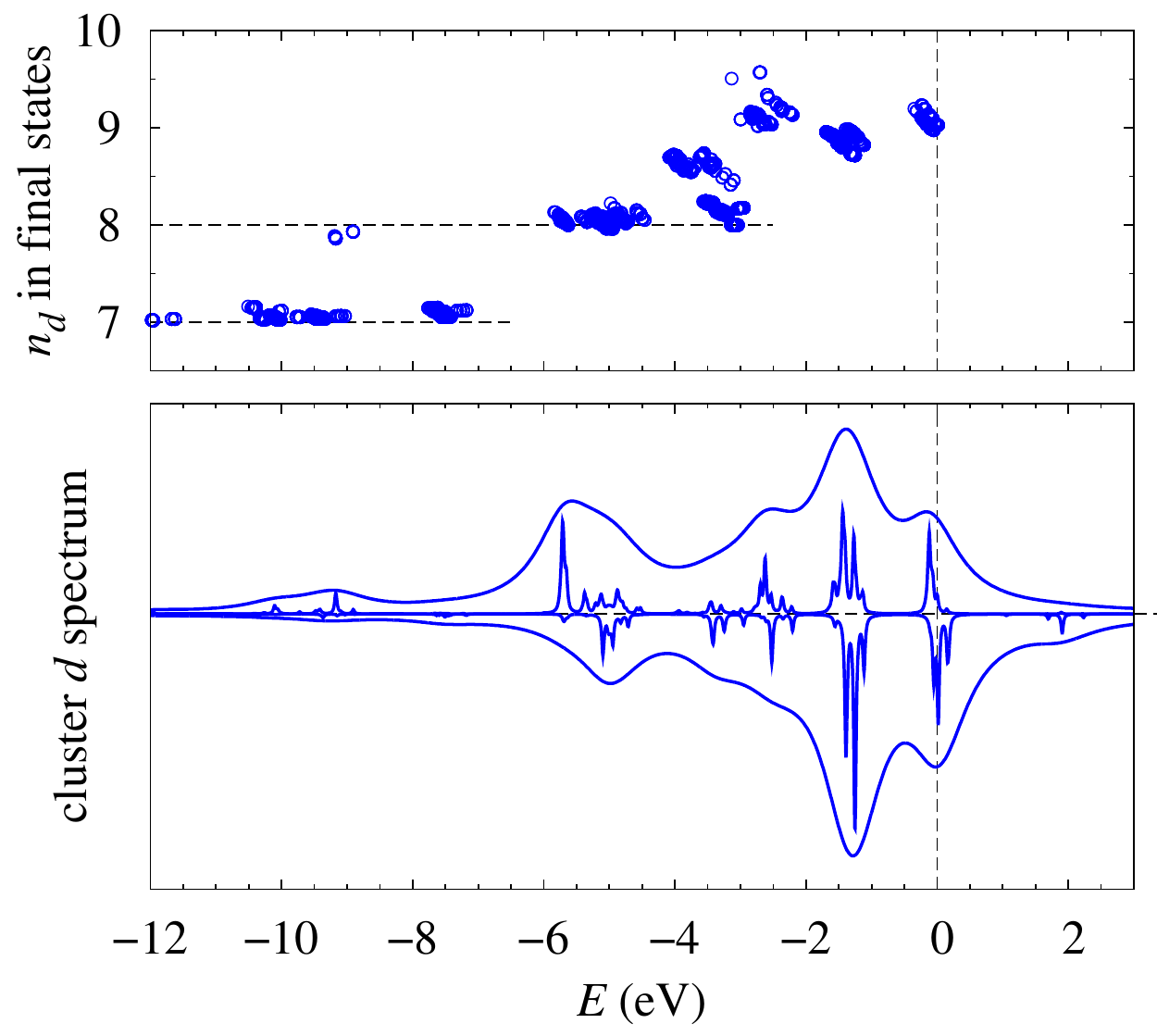}}
\caption{\label{fig:mag_cluster_spectrum}%
  (color online) Spectrum of the same model as
  Fig.~\ref{fig:TJS_spectrum}, but with the bath orbitals shifted
  to $-0.4$ eV. The ground state is magnetic in
  this case. Note the different range of the top panel compared to
  Fig.~\ref{fig:TJS_spectrum}.}
\end{figure}

It is instructive to compare the spectral functions corresponding
to the different cluster ground states. Figure~\ref{fig:TJS_spectrum}
shows the spectrum obtained when the bath orbitals are placed exactly at
the Fermi level, $\epsilon_{\rm b}=0$. The local moment induced by the
external magnetic field is negligible in this case and the spectral function is
nearly symmetric. The spectrum is practically identical to the result
presented in Ref.~\onlinecite{tanaka1992} as it should be, since we used
the same parameters. The
superimposed plot of the $d$-orbital filling $n_d$ in the
photoemission final states indicates that the satellite structures
around $-6.5$ eV and near $-10$ eV are due to the $d^8$ final states.


The spectral function corresponding to the bath orbitals lowered to
$-0.4$ eV is plotted in Fig.~\ref{fig:mag_cluster_spectrum}. The ``6 eV
satellite'' has now a shape similar to our DMFT solution
(Fig.~\ref{fig:dos_DMFT}) as well as to the experimental data: the
minority-spin component is less intense and is
located at smaller binding energies. Comparison of
Figs.~\ref{fig:TJS_spectrum} and~\ref{fig:mag_cluster_spectrum}
reveals that the composition of the final states constituting the main
$d$ band is shifted toward a larger average $n_d$, likely due to
an increased impurity--bath hybridization caused by the reduced
distance between the impurity and bath orbitals. Analogously, the
enhanced values $n_d\sim 8.5$ at the satellite in the DMFT solution
(Fig.~\ref{fig:dos_DMFT}) are
probably due to the hybridization with the extra shell of bath
orbitals not present in the model discussed in this appendix.

\bibliography{nickel,lanczos,dmft}

\end{document}